\documentclass{svmult}
\pdfoutput=1
\usepackage{graphicx}
\usepackage{multicol} 
\usepackage[bottom]{footmisc}

\usepackage{url}
\usepackage{epsfig}
\usepackage{graphicx}
\usepackage{tabularx}

\newcolumntype{Z}{>{\arraybackslash}X}
\newcolumntype{S}{>{\centering\small\arraybackslash}c} 
\begin{document}

\title*{Encoding changing country codes for the Semantic Web with ISO 3166 and SKOS}
\titlerunning{ISO 3166 country codes in SKOS}
\author{Jakob Vo\ss}
\institute{Verbundzentrale des GBV (VZG), 
Platz der G\"{o}ttinger Sieben 1,\\
37073 
G\"{o}ttingen, Germany.
\texttt{jakob.voss@gbv.de}}

\maketitle


\abstract{
This paper
shows how authority files can be encoded for the Semantic Web with the
Simple Knowledge Organisation System (SKOS). In particular the application of SKOS
for encoding the structure, management, and utilization of country codes as defined 
in ISO~3166 is demonstrated. The proposed encoding gives a use case for SKOS that 
includes features that have only been discussed little so far, such as multiple 
notations, nested concept schemes, changes by versioning.
}

\section{Introduction}

\subsection{Semantic Web}

The Semantic Web is a vision to extend the World Wide Web to a universal,
decentralised information space. To join in, information has to be expressed 
with the Resources Description Framework (RDF) in form of statements about 
resources. All resources are identified by Uniform Resource Identifiers (URIs) 
as defined in RFC~3986 \cite{RFC3986}.
URIs can identify documents, but also real-world objects and abstract 
concepts. In library and information science controlled vocabularies are used 
to uniformly identify objects --- also across different databases.
An example of such controlled vocabulary is
ISO~3166 \cite{ISO3166-1:2006} that defines codes and names to identify countries and their
subdivisions. To use ISO~3166 in Semantic Web applications for referring to countries, an 
encoding in RDF is needed. The encoding should
include explicit relations between codes in ISO~3166 and define a
way how to deal with changes. It is shown how the Simple
Knowledge Organisation Systems (SKOS) can be used to encode ISO~3166, and 
which parts of it need to be redefined to do so. Examples of RDF in this paper
are given in Notation~3 (N3) \cite{Notation3}.

\subsection{ISO~3166 and other systems of country codes}

Country codes are short codes that represent countries and dependent areas. The
most common code for general applications is ISO~3166, but there are many other
country codes for special uses. Country codes are managed by an agency that
defines a set of countries, with code, name and partly additional information. 
Examples of relevant systems of country codes beside ISO~3166 include 
codes that are used by the US government as defined by the Federal Information Processing Standard (FIPS), 
codes of the International Olympic Committee (IOC), 
codes of the World Meteorological Organization (WMO), 
and numerical country calling codes assigned by the International 
Telecommunications Union (ITU).
Some country codes occur as part of more general coding systems, for instance in the geographical table of Dewey Decimal Classification (DDC) that is used as a universal library
classification. Other systems also identify groups of countries such as the group 
identifiers of International Standard Book Numbers (ISBN).
More country code systems are listed in the English Wikipedia \cite{wiki:countrycodes}. The best public resource on country codes on the Web is Statoids \cite{Statoids} that includes references and a history of updated codes for many country subdivisions. GeoNames \cite{GeoNames} is an open, free-content geographical database that also contains countries and subdivisions. In contrast to ISO~3166 (which GeoNames partly refers to) GeoNames already uses URIs and SKOS to publish its content, but changes are rather uncontrolled because the database can be edited by anyone.
Examples of agencies that not define codes but names of countries and subdivisions are the
Board on Geographic Names (BGN) in the United States and the Permanent Committee on Geographical Names 
(StAGN) in Germany.

\subsection{ISO 3166}

ISO~3166 is an international standard for coding the names of countries and its
subdivisions. It consists of three parts. ISO~3166-1 (first published in 1974)
defines two letter codes, three letter codes and three digit numeric codes for
countries and dependent areas together with their names in English and French. 
The standard is widely refered to by other standards. For instance ISO~3166-2 is used
for most of the country code top-level domains as defined by Internet Assigned Numbers Authority (IANA) and the ICANN Country Code Names Supporting Organisation (ccNSO). ISO~3166-2 (first published 1998) builds on ISO~3166-1 and defines codes for country subdivisions. Figure~\ref{ISO3166:structure} shows the relations 
between ISO~3166, ISO~3166-1, and ISO~3166-2. ISO~3166-3 defines four letter codes for countries that merged, split up or changed the main part of their name and their two letter ISO~3166-1 codes
since 1974. ISO~3166 is continuously updated via newsletters that are 
published by the ISO~3166 Maintenance Agency.\footnote{\url{http://www.iso.org/iso/country_codes}} 
In November 2006 a second edition of ISO~3166-1 was published \cite{ISO3166-1:2006}. It contains a consolidation all changes to the lists of ISO~3166-1:1997, published in the ISO~3166 Newsletter up to V-12.
Meanwhile this edition has been corrected by a technical corrigendum
that was published in July 2007 \cite{ISO3166-1:2006Cor1}.

\subsection{SKOS}
SKOS was first developed in the SWAD-Europe project (2002-2004). It is a RDF-based standard for representing and sharing thesauri, classifications,
taxonomies, subject-heading systems, glossaries, and other controlled
vocabularies that are used for subject indexing in traditional Information
Retrieval. Examples of such systems are the AGROVOC Thesaurus, the Dewey Decimal Classification, 
and the dynamic category system of Wikipedia \cite{VossWikipediaTagging}.
Encoding controlled vocabularies with SKOS 
allows them to be passed between computer applications in an interoperable way and 
to be used in the Semantic Web. Because SKOS does not carry the strict 
and complex semantics of the Web Ontology Language (OWL),
it is also refered to as ``Semantic Web light''.
At the same time SKOS is compatible with OWL and can be extended with 
computational semantics for more complex applications.\cite{Sanchez-Alonso}
SKOS is currently being revised in the
Semantic Web Deployment Working Group of W3C to become a W3C Recommendation in 2008.

\section{Related Work}
Use cases and application guidelines for SKOS can best
be found at the SKOS homepage.\footnote{\url{http://www.w3.org/2004/02/skos/}}
Guidelines for using SKOS to encode thesauri \cite{SKOSThesauri,SKOSQuickThesauri} 
and classification schemes \cite{SKOSClassification} been 
published, while the use to encode authority files and standards like ISO~3166 
has not been analysed in detail so far. To a slightly lesser degree this also 
applies to revision and changes. Although changes are common in living Knowledge Organization Systems, research about this process is rare. The Fourth International Conference of the International Society for Knowledge Organization in 1996 \cite{ISKO2004} was about changes in general --- but the change only dealed about getting existing systems digital, a task that is still not finished and will hopefully bring more interoperability with SKOS. In computer science Johann Eder has done some recent work about modelling and detecting changes in ontologies \cite{ModellingChangesOntologies, DetectingChangesOntologies}. He presented an approach to represent changes in ontologies by introducing information about the valid time of concepts. Following this, a changed concept must get a new URI which is compatible to the method presented in this paper. Bakillah et al. \cite{bakillah2006} propose a semantic similarity model for multidimensional databases with different geospatial and temporal data -- however countries are more than simple, undisputed geographic objects. On the contrary is is unclear whether results from ontology evolution can be applied to knowledge organization systems. Noy and Klein\cite{OntologyEvolutionSchemaEvolution} argue that ontology versioning is different from schema evolution in a database -- the same applies to ontology versioning compared to changes in knowledge organization systems because the latter are mainly designed for subject indexing and retrieval without strict semantics and reasoning.

\section{Encoding ISO 3166 in SKOS}

\subsection{Basic elements}
The basic elements of SKOS are concepts (\texttt{skos:Concept}). A concept in
SKOS is a resource (identified by an URI) that can be used for subject indexing.
To state that a resource is indexed with a specific concept, SKOS provides the property
\texttt{skos:subject}. The concepts of ISO~3166 are countries and their subdivisions.
Hierarchical relations between concepts are encoded with
\texttt{skos:broader} and \texttt{skos:narrower}. These relationships allow
applications to retrieve resources that are index with a more specific
concept when searching for a general term \cite{MilesRetrieval}.
For representation and usage by humans, concepts are refered to by labels (names).
SKOS provides the labelling properties \texttt{skos:prefLabel} and
\texttt{skos:altLabel}. A concept should only have one \texttt{skos:prefLabel}
at least per language -- as shown below this causes problems due to 
the definition of `language'. The following example encodes basic 
parts of ISO~3166 for two concepts: France and the subordinated region Bretagne are encoded together with 
their English names and their ISO codes \texttt{FR} (`France') and \texttt{FR-E} (`Bretagne').
Unless the ISO~3166 Maintenance Agency defines an official URI schema, unspecified namespace prefixes like \texttt{iso3166:} are used:

\begin{small}
\begin{verbatim}
iso3166:FR a skos:Concept ;
  skos:prefLabel "France"@en ;
  skos:prefLabel "FR"@zxx ;
  skos:narrower iso3166:FR-E .

iso3166:FR-E a skos:Concept ;
  skos:prefLabel "Bretagne"@en ;
  skos:prefLabel "FR-E"@zxx ;
  skos:broader iso3166:FR-E .
\end{verbatim}
\end{small}

\subsection{Notations}
The main labels of ISO~3166 are not names but country codes. Such codes are 
also known as notations in other knowledge organisation systems. The final 
encoding method of notations in SKOS is still an open issue. The example
above uses ISO~639-2 language code \texttt{zxx} for `no linguistic content'
as proposed in \cite{SKOSClassification}.
This solution has some drawbacks: First the code 
was introduced the IANA language subtag registry in 2006, so not every
RDF application may already be aware of it. Second the SKOS specification 
requires the \texttt{skos:prefLabel} property to be unique per concept and 
language, so you can only specify one main notation per concept. The problem 
is caused by the special treatment of languages in RDF which is a 
failure by design\footnote{languages in RDF are not resources but absolute entities outside of RDF.}
To bypass the limitation, notations could either be implemented by additional labeling 
properties or by private language tags. If you use additional labeling properties 
for notations, SKOS must provide a way to state that a given property defines a notation.
This could be done with a new relation \texttt{skos:notationPropery}:

\begin{small}
\begin{verbatim}
iso3166: a skos:ConceptScheme ;
  skos:notationPropery iso3166:twoLetterCode ;
  skos:notationPropery iso3166:threeLetterCode ;
  skos:notationPropery iso3166:numericalCode .

iso3166:FR a skos:Concept ;
  skos:prefLabel "France"@en ;
  iso3166:twoLetterCode "FR" ;
  iso3166:threeLetterCode "FRA" ;
  iso3166:numericalCode "250" .
\end{verbatim}
\end{small}

\noindent With RFC~4646 \cite{RFC4646} you can now define private language tags in RDF.
These tags are seperated with the reserved single-character subtag `x'. This way you could
define the new language tag  \texttt{x-notation} for notations:

\begin{small}
\begin{verbatim}
iso3166:FR a skos:Concept ;
  skos:prefLabel "France"@en ;
  skos:prefLabel "FR@x-notation-twoletter" ;
  skos:prefLabel "FRA@x-notation-threeletter" ;
  skos:prefLabel "250@x-notation-numerical" .
\end{verbatim}
\end{small}

\noindent
Another advantage of private language codes is that you can use them 
at different levels, for instance \texttt{de-x-notation} for a German 
notation. No matter which solution will be used for encoding notations in SKOS, it has 
to be defined clearly in the SKOS standard or notations will not be usable
among different applications.

\subsection{Grouping}
\begin{figure}
\epsfig{file=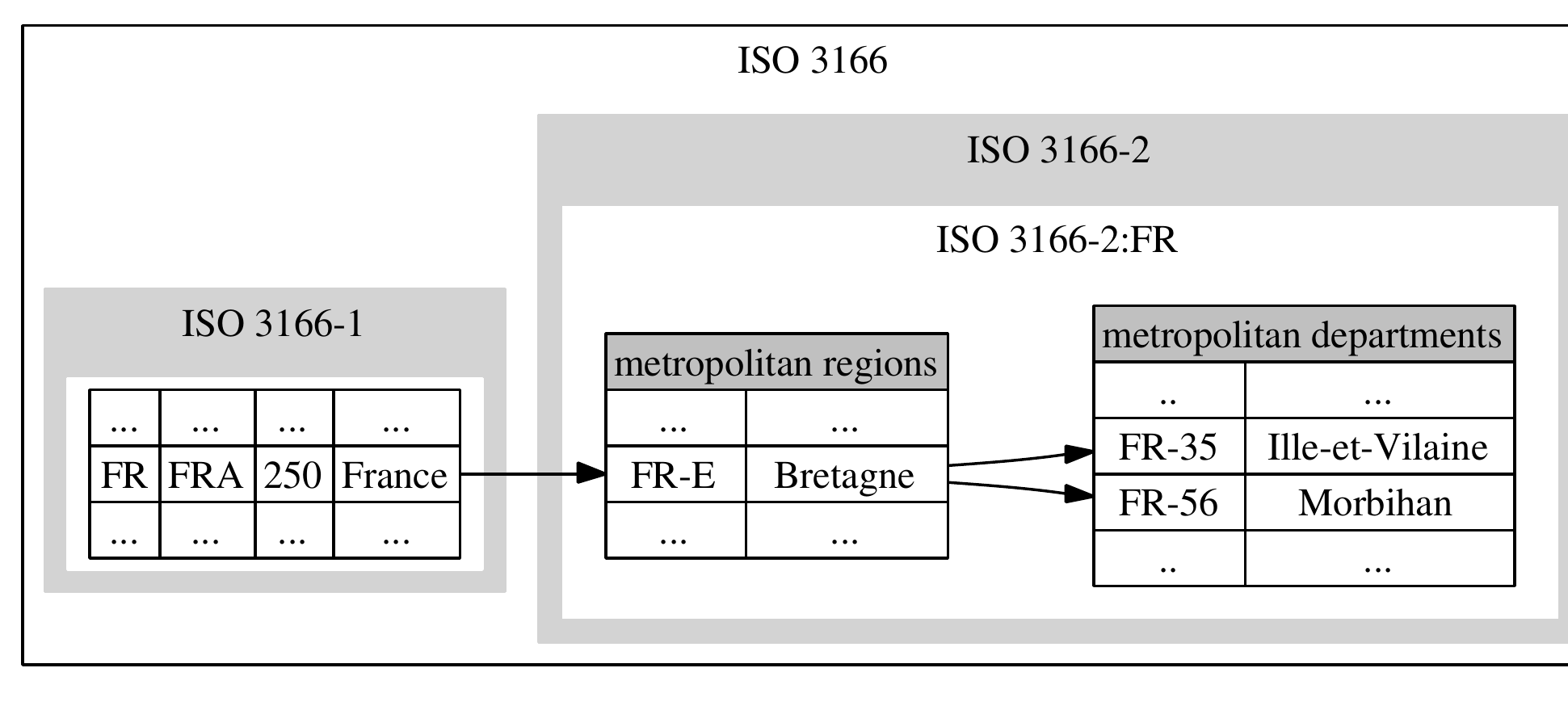, scale=0.6}
\caption{Internal structure and grouping of ISO~3166}
\label{ISO3166:structure}
\end{figure}

ISO~3166 is does not only consist of country codes but it also has an internal
structure. First the three parts ISO~3166-1, ISO~3166-2, and
ISO~3166-3 are concept schemes of their own but their concepts refer to each other.
Second the country subdivisions as defined in ISO~3166-2 can be grouped and build 
upon another. For instance France is divided
in 100 departments which are grouped into 22 metropolitan and four overseas
regions, and Canada is disjointedly composed of 10 provinces and 3 territories.
Figure~\ref{ISO3166:structure} shows the structure of ISO~3166 with an extract
of the definitions for France.

\noindent To encode groupings of concepts, SKOS provides the classes 
\texttt{skos:Collection} and \texttt{skos:ConceptScheme} and the properties
\texttt{skos:member} and \texttt{skos:inScheme}. The current standard
only allows \texttt{skos:Collection} to be nested. This is problematic 
for vocabularies like ISO~3166, that nested parts of which are also used 
independently. An easy solution is to make 
\texttt{skos:ConceptScheme} a subclass of \texttt{skos:Collection}.
This way concept schemes can be nested via \texttt{skos:member} 
(figure~\ref{ISO3166:structure-encoding}).

\begin{figure}
\begin{small}
\begin{verbatim}
iso3166: a skos:ConceptScheme ;
  skos:member iso3166-1: ;
  skos:member iso3166-2: .

iso3166-1: a skos:ConceptScheme .
iso3166-2: a skos:ConceptScheme ;
  skos:member iso3166-2:FR .

iso3166-2:FR a skos:ConceptScheme ;
  skos:member iso3166-2:FR-regions ;
  skos:member iso3166-2:FR-departements .

iso3166-2:FR-regions a skos:Collection ;
  skos:member iso3166:FR-E .

iso3166-2:FR-departements a skos:Collection ;
  skos:member iso3166:FR-35 ;
  skos:member iso3166:FR-56 .
\end{verbatim}
\end{small}
\caption{Proposed encoding of figure~\ref{ISO3166:structure} (without concepts)}
\label{ISO3166:structure-encoding}
\end{figure}

\subsection{Changes and versioning}

SKOS provides concept mapping relations to merge and combine identfiers 
from different concept schemes. A first working draft of the SKOS mapping 
vocabulary was published in 2004 \cite{skos_mapping}. It includes 
properties for concept equivalence (\texttt{skos:exactMatch}),
specialization (\texttt{skos:narrowMatch}), and 
concept generalization (\texttt{skos:broadMatch}).
In practise full one-to-one mappings between concept schemes
are rare because of differences in definition, focus, politics, and update
cycles. In the following it will be shown how mapping relations can be 
used to encode changes and versioning in ISO~3166. Mappings between 
different systems of country codes remains a topic to be analyzed 
in more detail. A promising candidate to start with for mapping to 
ISO~3166 would be the GeoNames database which already uses SKOS.\cite{GeoNames}

Nationalists might have a different opinion, but countries are no stable entities:
Contries come into existence, they can split and merge, change their names and area, or 
even disappear. To keep track of changes and the current situation, every 
modification in a schema of country codes needs to be documented for further 
lookup. The ISO~3166 Maintenance Agency uses newsletters and editions to publish 
updates. For Semantic Web applications these updates need to be explicitely 
specified in RDF. To develope a consistent encoding of changes, you must first 
consider all possible types of updates and paradigms of versioning. 
Types of changes are:

\begin{enumerate}
  \item A new country arises
  \item A country disappears
  \item A country is split into two or more countries
  \item Two or more countries unite (join)
  \item A country remains but its identity changes
\end{enumerate}

Type 1 and 2 are easy to model if there is no predecessor/successor but nowadays countries mostly arise from other countries (type 3 to 5). Easy examples of splits (type 3) are the division of Czechoslovakia 
(ISO code \texttt{CS}) into the Czech Republic (\texttt{CZ}) and Slovakia (\texttt{SK}) in 1993 and the division of Serbia and Montenegro (\texttt{CS}, until 2003 named Yugoslavia with code \texttt{YU}) into Serbia (\texttt{RS}) and Montenegro (\texttt{ME}) in 2006. An example of a simple join (type 4) is the German reunification in 1990. Other changes such as large reforms of country subdivisions and partly splits are more complex. They mostly imply that the identity of all involved entities change. To distinguish countries before and after a change, it is crucial to assigned a new URI for each version. The examples of Yugoslavia (which underwent several splits between 1991 and 2006) and the country code \texttt{CS} show that also controlled codes and names can be ambiguous if date is unknown and versioning is not respected.

You should keep in mind that changes in the basic structure of countries are political and 
can be highly  controversial. This means that the existence and nature of a change depends 
on who and when you ask. Encoding schemes of country codes can only give you guidance how to 
consistenly encode changes for reasoned retrieval but you first have to agree upon what 
happend with the involved entities.

The encoding of changes in ISO~3166 in SKOS will be shown with the example of Canada. Canada, 
the world second largest country in total area, is composed of 
10 provinces and 3 territories. The provinces are independent states with 
own jurisdiction. In March 31, 1949 Newfoundland entered the Canadian 
confederation as the 10th province. The territories cover the parts 
of Canada that do not belong to provinces. They are created by the 
federal government and have less authority. The North-Western Territory was 
formerly much larger then today. It contained parts of current provinces and 
the area that now form the territories Yukon (since 1898) and Nunavut (1999).
Between 1998 and 2002 the ISO 3166-2 entry of Canada has been changed three 
times. Figure~\ref{tab:changes} contains an overview of the changes:

\begin{itemize}
\item Newsletter I-1 (2000-06-21) Addition of 1 new territory: 
The new territory Nunavut split up from Northwest Territories.
\item Newsletter I-2 (2002-05-21) Correction of name form of CA-NF: 
The name 'Newfoundland' changed to 'Newfoundland and Labrador'.
\item Newsletter I-4 (2002-12-10) Change of code element of Newfoundland 
and Labrador: The country code \texttt{CA-NF} changed to \texttt{CA-NL}.
\end{itemize}

\begin{figure}
\begin{tabular}{|S|S|S|S|S|}
\hline
{\bfseries Initial} & {\bfseries Newsletter} & {\bfseries Newsletter} 
                    & {\bfseries Newsletter} & {\bfseries Newsletter} \\
{\bfseries Version} & {\bfseries I-1} & {\bfseries I-2} 
                    & {\bfseries I-3} & {\bfseries I-4} \\
\hline\hline
\multicolumn{2}{|c|}{ \texttt{CA-NF} } & 
\multicolumn{2}{c|}{ \texttt{CA-NF} } & 
\texttt{CA-NL}
\\
\multicolumn{2}{|c|}{ Newfoundland } &
\multicolumn{2}{c|}{ Newfoundland } &
Newfoundland
\\
\multicolumn{2}{|c|}{ } &
\multicolumn{2}{c|}{ and Labrador } &
and Labrador
\\
\hline
\texttt{CA-NT} & \multicolumn{4}{c|}{ \texttt{CA-NT} } \\
Northwest & \multicolumn{4}{c|}{ Northwest Territories } \\\cline{2-5}
Territories & \multicolumn{4}{c|}{ \texttt{CA-NU} } \\
& \multicolumn{4}{c|}{ Nunavut } \\
\hline
\end{tabular}
\caption{Changes of Canada in ISO 3166-2}
\label{tab:changes}
\end{figure}

To model these changes, unique URIs must be defined for each version --
at least when the definition of a country or country subdivision changed.
For easy detection of the valid URI for a given date or newsletter, 
a directory structure of URLs with namespaces for each newsletter 
should be provided by the ISO~3166 Maintenance Agency.
Changing country codes are then mapped to each other with the SKOS Mapping vocabulary. 
For codes that did not change with a newsletter, you could either provide 
new URIs and connect unmodified 
concepts with the \texttt{owl:sameAs} property from the OWL Web Ontology Language
or just direct to the previous URI with a HTTP 303 redirect. Support of any method 
in SKOS applications can be ensured by best practise rules in the final SKOS 
standards.
Figure~\ref{fig:changesenc} contains an encoding of the changes of
Canada in ISO~3166 as shown in figure~\ref{tab:changes}. The change 
of Newfoundland to Newfoundland and Labrador in newsletter I-2 and I-4
is encoded by an exact mapping between sequent versions 
(\texttt{skos:exactMatch}) while the split of Northwest Territories 
in newsletter I-1 is encoded by an \texttt{skos:narrowMatch}. Unchanged
country codes are connected with \texttt{owl:sameAs}.

\begin{figure}
\begin{small}
\begin{verbatim}
@prefix iso3166-2-v0: <http://iso.org/iso3166/2/first/> .
@prefix iso3166-2-v1: <http://iso.org/iso3166/2/newsletter-1/> .
@prefix iso3166-2-v2: <http://iso.org/iso3166/2/newsletter-2/> .
@prefix iso3166-2-v3: <http://iso.org/iso3166/2/newsletter-3/> .
@prefix iso3166-2-v4: <http://iso.org/iso3166/2/newsletter-4/> .
@prefix iso3166-2-v4: <http://iso.org/iso3166/2/current/> .

iso3166-2-v0:CA-NF  owl:sameAs        iso3166-2-v1:CA-NF .
iso3166-2-v0:CA-NT  skos:narrowMatch  iso3166-2-v1:CA-NT .
iso3166-2-v0:CA-NT  skos:narrowMatch  iso3166-2-v1:CA-NU .

iso3166-2-v1:CA-NF  skos:exactMatch   iso3166-2-v2:CA-NF .
iso3166-2-v1:CA-NT  owl:sameAs        iso3166-2-v2:CA-NT .
iso3166-2-v1:CA-NU  owl:sameAs        iso3166-2-v2:CA-NU .

iso3166-2-v2:CA-NF  owl:sameAs        iso3166-2-v3:CA-NF .
iso3166-2-v2:CA-NT  owl:sameAs        iso3166-2-v3:CA-NT .
iso3166-2-v2:CA-NU  owl:sameAs        iso3166-2-v3:CA-NU .

iso3166-2-v3:CA-NF  skos:exactMatch   iso3166-2-v4:CA-NL .
iso3166-2-v3:CA-NT  owl:sameAs        iso3166-2-v4:CA-NT .
iso3166-2-v3:CA-NU  owl:sameAs        iso3166-2-v4:CA-NU .
\end{verbatim}
\end{small}
\caption{Encoding of changes of Canada in ISO 3166-2}
\label{fig:changesenc}
\end{figure}

\section{Summary and Conclusions}
With the Simple Knowledge Organisation System more and more thesauri, classifications, subject-heading systems, and other controlled vocabularies can be integrated into the Semantic Web.
This will increase interoperability among Knowledge Organisation Systems 
which are already used and maintained for a long time and in many applications.
One kind of Knowledge Organisation Systems are Country codes, a common type of
authority files. This paper shows how in particular country 
codes from ISO~3166 can be encoded in RDF with SKOS. 
ISO~3166 and its parts are widely used and referred to by other applications 
and standards that could benefit from such a common encoding.
ISO~3166 includes some particular features of controlled vocabularities that have 
not been discussed in detail so far in the context of SKOS. 
The hereby proposed encoding contains support of country names and codes 
(notations), internal structure and nested concept schemes (grouping), and 
versioning of changes. To explicitly support notations a notation property
or a private language subtag (\texttt{x-notation}) has to be defined. Nested
concept schemes can easily be supported by making \texttt{skos:ConceptScheme} 
a subclass of \texttt{skos:Collection}. Finally you can track changes by
publishing new URIs for the concepts of each version of a concept scheme 
and interlink them with \texttt{owl:sameAs} and SKOS mapping relations.

To get a reliable RDF representation of ISO~3166, that other Semantic Web applications can 
build upon, the upcoming W3C Recommendation of SKOS must first be finalized with support of notations, grouping concept schemes and versioning. Second an URL scheme for country codes of ISO~3166 has to be defined by ISO, and third the ISO~3166 Maintenance Agency must regularly and freely publish versioned ISO~3166 data in SKOS. A public, official, RDF-representation of ISO~3166 will allow heterogeneous data on the web to be linked for homogeneous, semantic retrieval via aggregating resources. For instance 
statistics by the United Nations can be combined with encyclopaedic information by Wikipedia and 
visualised with geographical data by GeoNames. With controlled versioning and linking to specific 
versions you can also access historic information without having to update all involved datasets.
Geographic data from GeoNames could be used to select a country or country subdivision 
by browsing on a map. Linked with ISO~3166 in SKOS then relevant past countries could be determined 
to extend searches in databases with other country codes.
In this way ISO~3166 and other authority files will be the corner stones of connecting distributed data to a universal, decentralised information space.

\vfill
\noindent
This paper is accepted to appear in the proceedings of the 2nd International 
Conference on Metadata and Semantics Research (MTSR 2007), published by Springer.

\pagebreak


\begin{thebibliography}{10}

\bibitem{RFC3986}
Berners-Lee, T., Fielding, R., Masinter, L.:
\newblock {Uniform Resource Identifier (URI): Generic Syntax}.
\newblock RFC 3986 (January 2005)

\bibitem{ISO3166-1:2006}
{International Organization for Standardization}:
\newblock {Codes for the representation of names of countries and their
  subdivisions --- Part~1: Country codes}.
\newblock ISO3166:1-2006 (November 2006)

\bibitem{Notation3}
Berners-Lee, T.:
\newblock {Notation3 (N3) A readable RDF syntax}

\bibitem{wiki:countrycodes}
{Wikipedia}:
\newblock {Country code}.
\newblock \url{http://en.wikipedia.org/w/index.php?oldid=154259104} (August
  2007)

\bibitem{Statoids}
Law, G.:
\newblock Statoids.
\newblock \url{http://www.statoids.com} (2007) [visited 2008-01-20].

\bibitem{GeoNames}
Wick, M.:
\newblock Geonames.
\newblock \url{http://www.geonames.org} (2007) [visited 2008-01-20].

\bibitem{ISO3166-1:2006Cor1}
{International Organization for Standardization}:
\newblock {ISO 3166-1 Technical Corrigendum~1}.
\newblock ISO3166:1-2006/Cor 1:2007 (July 2007)

\bibitem{VossWikipediaTagging}
Voss, J.:
\newblock {Collaborative thesaurus tagging the Wikipedia way}.
\newblock \url{http://arxiv.org/abs/cs/0604036v2} (April 2006)

\bibitem{Sanchez-Alonso}
Sanchez-Alonso, S., Garcia-Barriocanal, E.:
\newblock {Making use of upper ontologies to foster interoperability between
  SKOS concept schemes}.
\newblock Online Information Review \textbf{30}(3) (May 2006)  263--277

\bibitem{SKOSThesauri}
van Assem, M., Malaisé, V., Miles, A., Schreiber, G.:
\newblock {A Method to Convert Thesauri to SKOS}.
\newblock The Semantic Web: Research and Applications  95--109

\bibitem{SKOSQuickThesauri}
Miles, A.:
\newblock {Quick Guide to Publishing a Thesaurus on the Semantic Web}.
\newblock \url{http://www.w3.org/TR/swbp-thesaurus-pubguide} (May 2005)

\bibitem{SKOSClassification}
Voss, J.:
\newblock {Quick Guide to Publishing a Classification Scheme on the Semantic
  Web}.
\newblock \url{http://esw.w3.org/topic/SkosDev/ClassificationPubGuide}
  (September 2006) [last changed 2006-09-11 13:08:41].

\bibitem{ISKO2004}
{Knowledge organization and change: Proceedings of the Fourth International
  ISKO Conference }, Indeks (1996)

\bibitem{ModellingChangesOntologies}
Eder, J., Koncilia, C.:
\newblock {Modelling Changes in Ontologies}.
\newblock Lecture Notes in Computer Science 3292 (2004)  662--673

\bibitem{DetectingChangesOntologies}
Eder, J., Wiggisser, K.:
\newblock {Detecting Changes in Ontologies via DAG Comparison}.
\newblock Lecture Notes in Computer Science 4495 (2007)  21--35

\bibitem{bakillah2006}
Bakillah, Mostafavi, B{\'e}dard:
\newblock {A Semantic Similarity Model for Mapping Between Evolving Geospatial
  Data Cubes}.
\newblock On the Move to Meaningful Internet Systems 2006: OTM 2006 Workshops
  (2006)

\bibitem{OntologyEvolutionSchemaEvolution}
Noy, N., Klein, M.:
\newblock {Ontology Evolution: Not the Same as Schema Evolution}.
\newblock Knowledge and Information Systems (2003)

\bibitem{MilesRetrieval}
Miles, A.:
\newblock {Retrieval and the Semantic Web}.
\newblock \url{http://isegserv.itd.rl.ac.uk/retrieval/} (September 2006)

\bibitem{RFC4646}
Phillips, Davis:
\newblock {Tags for the Identification of Languages}.
\newblock RFC 4646 (September 2006)

\bibitem{skos_mapping}
Miles, A., Brickley, D.:
\newblock {SKOS Mapping Vocabulary Specification}.
\newblock \url{http://www.w3.org/2004/02/skos/mapping/spec/} (November 2004)

\end{thebibliography}

\end{document}